\documentclass[preprint2,10.5pt]{aastex}
\begin{document}
\title{\large{\rm{ANCHORING THE DISTANCE SCALE VIA X-RAY/IR DATA \\ FOR CEPHEID CLUSTERS: SU CAS}}}
\author{D. Majaess$^1$, D. G. Turner$^1$, L. C. Gallo$^1$, W. Gieren$^2$, C. Bonatto$^3$ \\ D. J. Lane$^1$, D. Balam$^4$, L. Berdnikov$^{5,6}$}
\affil{$^1$Saint Mary's University, Halifax, NS, Canada.}
\affil{$^2$Universidad de Concepci\'on, Concepci\'on, Chile.}
\affil{$^3$Universidade Federal do Rio Grande do Sul, Porto Alegre, RS, Brazil.}
\affil{$^4$Dominion Astrophysical Observatory, Victoria, BC, Canada.}
\affil{$^5$Moscow M V Lomonosov State University, Sternberg Astronomical Institute, Moscow, Russia.}
\affil{$^6$Isaac Newton Institute of Chile, Moscow Branch, Moscow, Russia.}
\email{dmajaess@cygnus.smu.ca}

\begin{abstract}
New X-ray (XMM-Newton) and $JHK_s$ (OMM) observations for members of the star cluster Alessi 95, which \citet{tu12} discovered hosts the classical Cepheid SU Cas, were used in tandem with UCAC3 (proper motion) and 2MASS observations to determine precise cluster parameters: $E(J-H)=0.08\pm0.02$ and $d=405\pm15$ pc.  The ensuing consensus among cluster, pulsation, IUE, and trigonometric distances ($d=414\pm 5(\sigma_{\bar{x}}) \pm 10 (\sigma )$ pc) places SU Cas in a select group of nearby fundamental Cepheid calibrators ($\delta$ Cep, $\zeta$ Gem).  High-resolution X-ray observations may be employed to expand that sample as the data proved pertinent for identifying numerous stars associated with SU Cas. Acquiring X-ray observations of additional fields may foster efforts to refine Cepheid calibrations used to constrain $H_0$.
 \end{abstract}


\section{{\rm \footnotesize INTRODUCTION}}
\citet{tu12} discovered that the $1.95^{\rm d}$ classical Cepheid SU Cas is a member of the star cluster Alessi 95.  The discovery permitted the fundamental parameters ($\log{\tau}, {\cal M_{*}/M_{\sun}},E_{B-V}, M_V , R_{*}/R_{\sun}$) for SU Cas to be inferred from cluster membership, thereby enabling a long-standing debate concerning the Cepheid's distance and pulsation mode to be resolved \citep[][and references therein]{gi76,gi82,ev91,us01,tu12}.  Previous uncertainties associated with SU Cas permeated into efforts to anchor the short-period regime of the Cepheid period-luminosity relation.  SU Cas is the vital link since the variable is among the shortest-period Galactic Cepheids known \citep{be08}, and the nearest such star \citep{be00}\footnote{Polaris is nearer \citep{tu09,tu12}, yet exhibits a pulsation period $\sim2\times$ larger than SU Cas \citep{be08}.  However, that does not mitigate Polaris' importance for the distance scale.}.  The Cepheid's proximity allows for independent confirmation of its distance via trigonometric parallax \citep[Hipparcos,][]{pe97,vl07,tu12}.  A principal research objective is to establish Cepheid calibrators with distances secured by independent means: e.g., infrared surface brightness technique \citep[ISB,][]{fg97,gi05,ba09,st11}, cluster membership \citep{tu10,ma11b}, IUE-binary \citep{ev92,ev95}, and trigonometric parallaxes \citep{vl07,be07}.  Achievement of that goal will facilitate efforts by the Carnegie Hubble \citep{fm10} and S$H_0$ES \citep{mr09} projects to determine $H_0$ reliably, and break degeneracies complicating the selection of a cosmological model \citep{ri11}.  Incidentally, the prototype of the class ($\delta$ Cep) has the most precise Cepheid distance established ($d=272\pm 3(\sigma_{\bar{x}}) \pm 5 (\sigma )$ pc), which is tied to a cluster distance from $UBVJHK_s$ photometry \citep{ma12}, Hipparcos trigonometric parallaxes for cluster stars \citep{ze99,vl07,ma12}, an ISB distance \citep{gi05,st11}, and Hipparcos/HST trigonometric parallaxes for the Cepheid and its companion HD213307\footnote{XMM-Newton data confirm that HD213307 features a companion \citep{be02}.  $\delta$ Cep is thus a multiple system, somewhat analogous to Polaris \citep{ev10}.} \citep{be02,vl07}.  $\zeta$ Gem is another nearby Cepheid which exhibits a precise mean distance \citep[$d=363\pm 9(\sigma_{\bar{x}}) \pm 26 (\sigma )$ pc,][]{ma12b} tabulated from cluster membership and HST/Hipparcos parallaxes.  The desire to incorporate SU Cas into that select sample of fundamental Cepheid calibrators is the impetus for this research \citep[see also][]{tu12}.

In this study, new X-ray and $JHK_s$ data from XMM-Newton \citep{ja01} and l'Observatoire Mont-M\'{e}gantique \citep[OMM,][]{ar10} are presented for Alessi 95.  X-ray observations provide a means of segregating cluster members from field stars \citep{ra96,ev11}, while OMM data supply more reliable photometry for fainter stars than 2MASS.

\begin{figure}[!t]
\epsscale{0.8}
\plotone{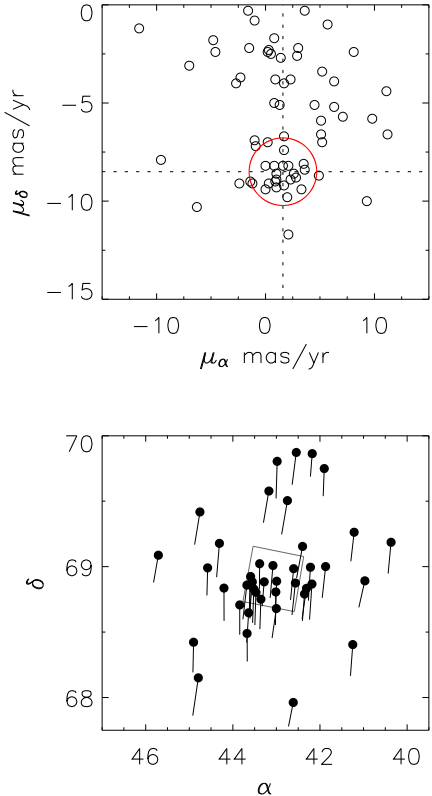}
\caption{\small{Top panel, UCAC3 proper motions for brighter stars within $r\sim35 \arcmin$ of SU Cas.  Cluster members aggregate near the proper motion estimate for SU Cas ($\mu_{\alpha}=1.0\pm0.7$ and $\mu_{\delta}=-8.6\pm0.8$ mas/yr).  Field contamination becomes more acute with increasing distance from the cluster center, thus the sample was restricted to $r\le35\arcmin$ to highlight the cluster's significance.  Bottom panel, the cluster core is discernible in the spatial distribution of stars $r\le60 \arcmin$ from SU Cas which exhibit $\mu_{\alpha}=-0.5\rightarrow5.0$ and $\mu_{\delta}=-6.0\rightarrow-10.5$ mas/yr ($\sigma<5$ mas/yr).  Proper motion trajectories ($10^5$ yr) are overplotted merely for illustrative purposes.  The area bounded by the solid lines represents the approximate region sampled by XMM-Newton.}}
\label{fig-pm}
\end{figure}

\begin{figure}[!t]
\epsscale{0.9}
\plotone{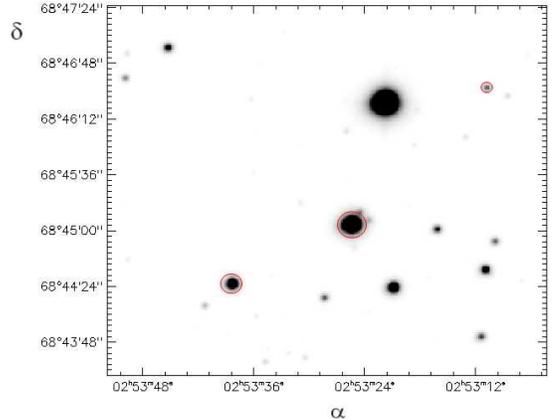}
\caption{\small{Finder chart (OMM $J$) for three probable cluster members which are X-ray sources.  The two brighter objects exhibit UCAC3 proper motions (e.g., $\mu_{\alpha}=0.8\pm1.0$ and $\mu_{\delta}=-8.2\pm0.8$ mas/yr) matching that established for SU Cas ($\mu_{\alpha}=1.0\pm0.7$ and $\mu_{\delta}=-8.6\pm0.8$ mas/yr).  The stars are probable cluster members based on X-ray, proper motion, and color-magnitude data (Fig.~\ref{fig-cmd}).   The finder chart represents a small portion of the observed field.}}
\label{fig-chart}
\end{figure}

\begin{figure*}[!t]
\epsscale{2.1}
\plotone{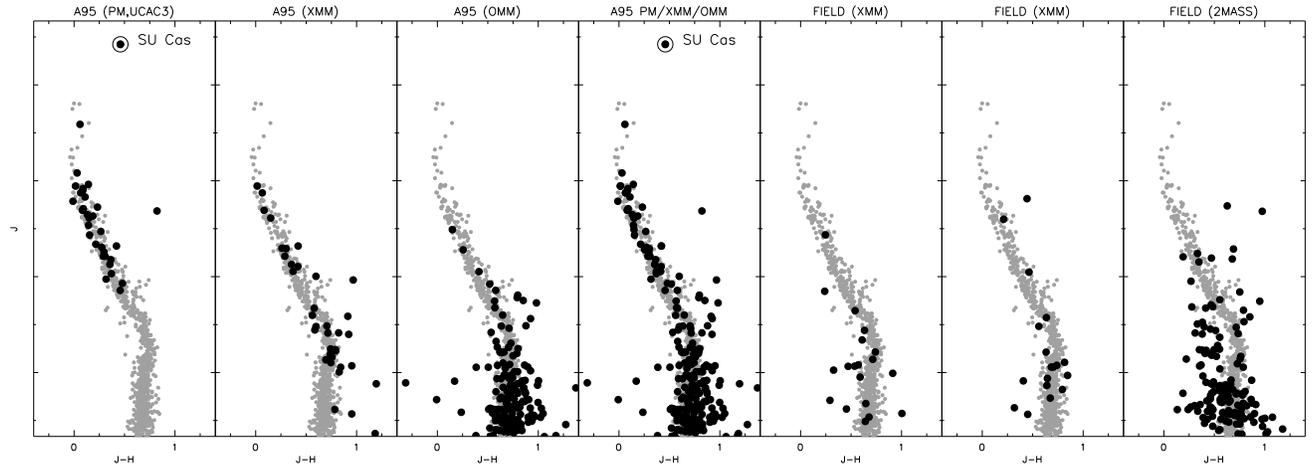}
\caption{\small{$JH$ color-magnitude diagrams for stars exhibiting similar proper motions to SU Cas (PM, UCAC3), stars emitting X-rays (XMM), and OMM stars sampled $r<4 \arcmin$ from the center of Alessi 95 (A95).  Gray dots are calibration stars from \citet{ma11}.  Comparison fields sampled by XMM-Newton and 2MASS imply that the cluster (A95) distribution is not fortuitous.}}
\label{fig-cmd}
\end{figure*}

\section{{\rm \footnotesize ANALYSIS}}
\subsection{{\rm \footnotesize BRIGHT CLUSTER MEMBERS}}
The positions for stars $r<35 \arcmin$ from SU Cas that feature UCAC3 \citep[Third U.S. Naval Observatory CCD Astrograph Catalog,][]{za10} proper motions and 2MASS photometry are shown in Fig.~\ref{fig-pm} (top panel).  The stellar overdensity near the proper motion for SU Cas ($\mu_{\alpha}=1.0\pm0.7$ and $\mu_{\delta}=-8.6\pm0.8$ mas/yr) represents the cluster Alessi 95.  The analysis was subsequently expanded and forty-five stars with similar proper motions to SU Cas ($\mu_{\alpha}=-0.5\rightarrow5.0$ and $\mu_{\delta}=-6.0\rightarrow-10.5$ mas/yr, $\sigma<5$ mas/yr) were detected within $\sim 1 \degr$ of the Cepheid.  The positions and proper motions of those objects are displayed in Fig.~\ref{fig-pm} (bottom panel), and the cluster core is discernible in that diagram.   Twenty-seven of the proper motion selected stars within $r\la40 \arcmin$ of SU Cas (Fig.~\ref{fig-pm}) form a distinct main-sequence in the $JH$ color-magnitude diagram (Fig.~\ref{fig-cmd}).  Only one star from the proper motion selected sample (Fig.~\ref{fig-pm}) exhibiting $r\la40 \arcmin$ appears to be a non-member, reaffirming that the proper motion data efficiently separate cluster members from field stars (Figs.~\ref{fig-pm},~\ref{fig-cmd}).  An inability to distinguish cluster members from field stars hinders isochrone fitting, and exacerbates uncertainties tied to the derived distance.  For example, the host cluster for $\zeta$ Gem proved more difficult to assess owing to the small offset of its proper motion relative to the field, particularly given the uncertainties \citep{ma12b}.  That shortcoming may be addressed by acquiring XMM-Newton/Chandra observations to identify cluster members, as later-type stars associated with $\zeta$ Gem \citep[$\log{\tau}=7.85\pm0.15$,][]{ma12} may emit X-rays owing to their comparative youth \citep{ra96,ev10,ev11}.  Independent confirmation of the $\zeta$ Gem cluster \citep{ma12} is desirable. 

X-ray data for the SU Cas field (PI Guinan) were obtained from the XMM-Newton public data archive (XSA).  Point sources identified via the XMM-Newton (PPS) Processing Pipeline Subsystem (Fig.~\ref{fig-chart}) were correlated with 2MASS photometry (closest source within $0.1\arcmin$).  The majority of the X-ray sources lie directly on the cluster main-sequence (Fig.~\ref{fig-cmd}), and the field stars are readily discernible in the diagram (e.g., objects redder than M-type cluster dwarfs).  X-ray observations are a pertinent means for identifying cluster members in harmony with other methods.  The brightest star in Fig.~\ref{fig-chart} \citep[BD+68$\degr$201, B9III-IV,][]{tu12} may exhibit a later-type companion which is the source of the X-ray emission, since later-type B stars are typically X-ray quiet \citep[][and references therein]{ev11}.  BD+68$\degr$201 and SU Cas exhibit analogous UCAC3 proper motions, to within the uncertainties. The faintest star in Fig.~\ref{fig-chart} is likely an M-type cluster dwarf, as implied by its $JHK_s$ photometry.  A list of the X-ray sources and their corresponding 2MASS data are available via the XMM-Newton public data archive (PPS products). 

\subsubsection{{\rm \footnotesize MEAN REDDENING FOR ALESSI 95}}
A $JHK_s$ color-color analysis of the UCAC3/XMM sample was performed to determine the mean reddening.  The $JHK_s$ intrinsic relations of \citet{sl09} and \citet{tu11} were employed to infer a mean color excess of $E(J-H)=0.08\pm0.02$.  That result is tied to a reddening law of $E(J-H)/E(H-K_s)=2.0$ \citep{sl09,ma11b}.  Stars which lie above the main-sequence (potential binary systems) in the color-magnitude diagram were excluded from the determination, in tandem with other stars exhibiting anomalous $JHK_s$ colors.  The mean color excess determined agrees with the optical excess established by \citet{lc07}, \citet{ko08}, and \citet{tu12} \citep[see also Table~5 in][and references therein]{ev91}. The conversion from the infrared to optical reddening is somewhat uncertain \citep[][and references therein]{bo04,ma08}.  Therefore, the optical reddening ($E_{B-V}\sim0.30\pm0.05$) derived from $E(J-H)$ exhibits larger uncertainty to reflect that ambiguity.

\subsection{{\rm \footnotesize FAINTER CLUSTER MEMBERS}}
Deeper $JHK_s$ images were obtained for the SU Cas field from the OMM.  The OMM houses a 1.6-m telescope equipped with a near-infrared wide-field imager (CPAPIR\footnote{\url{http://www.astro.umontreal.ca/cpapir}}).  The images were acquired on January 11, 2012.  PSF photometry was performed using DAOPHOT \citep{st87}.  The instrumental photometry was subsequently tied to 2MASS secondary standards in the field.  A color-magnitude diagram was tabulated for stars within $r<4 \arcmin$ of the cluster center to mitigate field contamination.  OMM $JHK_s$ photometry for stars near the core of Alessi 95 shall be tabulated online at CDS or WEBDA \citep{pa08}.  

Comparison fields displayed in Fig.~\ref{fig-cmd} reaffirm the existence of Alessi 95.  However, the analysis is complicated by the cluster's extent, high latitude position ($\ell,b\sim133,9\degr$) and inhomogeneous reddening.  The first XMM-Newton comparison field lies beyond the coronal radius (02:05:55 +64:56:33, $r\sim6 \degr$), however, hierarchical clustering observed in giant molecular clouds implies that members may still be sampled. The second XMM-Newton comparison field encompasses another classical Cepheid ($\beta$ Dor).  The 2MASS comparison field (Fig.~\ref{fig-cmd}, far right panel) is equal in area to the $r\sim4 \arcmin$ OMM sampling, but traces a symmetric annulus $r\sim5 \degr$ away. Field contamination will be most acute for the fainter OMM data.

No low-mass stars are visible in HST WFC3 images for SU Cas \citep[HST Proposal 12215,][]{ev09}. The saturation from the Cepheid was treated by subtracting (normalized) the image from a master, which was constructed (median combine) using numerous Cepheids observed for proposal 12215 \citep[see also][]{ev11}.  The late B-type (B9.5V) companion discovered by \citet[][her Fig.~4]{ev91} via IUE observations was not detected, indicating that the star is in rather close proximity to the Cepheid.  A  comprehensive analysis is forthcoming (Evans et al., in prep.).

\subsection{{\rm \footnotesize DISTANCE TO ALESSI 95}}
A precise distance may be established for the cluster via main-sequence fitting.  The infrared (2MASS/NOMAD) calibration presented by \citet{ma11} was employed since it offers numerous advantages.  First, the $JHK_s$ calibration is comparatively insensitive to variations in [Fe/H] and age \citep[][see also \citealt{sl09}]{ma11}.  Second, the deep $JHK_s$ photometry extends into the low mass regime ($\simeq0.4 M_{\sun}$) where $J-K_s$ remains constant with increasing magnitude ($M_J\gtrsim6$) for low mass M-type dwarfs, and $J-H$ exhibits an inversion \citep[see Fig.~\ref{fig-cmd},][and references therein]{ma11}.  The trends ensure precise \textit{JHK$_s$} main-sequence fits by providing distinct anchor points in color-magnitude and color-color diagrams (Fig.~\ref{fig-cmd}).  Third, the calibration's zero-point is tied to seven benchmark open clusters ($d<250$ pc) which exhibit matching $JHK_s$ and revised Hipparcos distances \citep[the Hyades, $\alpha$ Per, Praesepe, Coma Ber, IC 2391, IC 2609, and NGC 2451,][see also \citealt{mc11}]{vl09,ma11}.  The scale is anchored using clusters where consensus exists, rather than the discrepant case (i.e. the Pleiades).   The objective is to avoid deriving distances to Cepheid clusters using a single benchmark cluster, and prevent the propagation of ambiguity into the Cepheid calibration.  The revised Hipparcos distance for the Pleiades is $d=120.2\pm1.9$ pc \citep{vl09}, whereas $JHK_s$ data implied $d=138\pm6$ pc \citep{ma11}, and \citet{so05} employed HST to deduce $d=134.6\pm3.1$ pc.  By contrast, \citet{vl09} obtained $d=172.4\pm2.7$ pc for $\alpha$ Per, which matches that established via the $JHK_s$ analysis \citep[][their Table~1]{ma11}.  Fourth, potential variations in the $JHK_s$ reddening and extinction laws are predicted to be comparatively smaller than in the optical.  Obscuration by dust is less significant in the infrared \citep[$E_{J-H}\sim0.3 \times E_{B-V}$,][and references therein]{ma08,bo04}, which consequently mitigates the impact of variations in $R_{\lambda}$ ($J_0=J-E_{J-H} \times R_J$).  The ratio of total to selective extinction $R_J$ was adopted from \citet{ma11b} \citep[see also][]{bo04}.  The offset from the calibration yields a distance to Alessi 95 of $d=405\pm15$ pc.

\subsection{{\rm \footnotesize DISTANCE TO SU CAS}}
\label{s-dsucas}
A mean distance to SU Cas may be derived from the new $JHK_s$ cluster parameters established here (the Cepheid lies near cluster center), and the following estimates: the distance inferred from $UBV$ photometry and optical spectra for cluster members \citep[$d=429\pm8$ pc,][]{tu12}, the mean Hipparcos parallax for cluster stars \citep[$420\pm33$ pc,][]{vl07,tu12}, the ISB distance for SU Cas \citep[$d=414\pm12$ pc,][]{st11,tu12}, the original Hipparcos Cepheid parallax \citep[$d=433\pm116$ pc,][]{pe97}, and the revised Hipparcos Cepheid parallax \citep[$d=395\pm50$ pc,][]{vl07}.  A weighted mean of $d=414\pm 5(\sigma_{\bar{x}}) \pm 10 (\sigma )$ pc was established after assigning $w=3$ to the ISB, $JHK_s$, and $UBV$ distances, $w=2$ for the revised Hipparcos distance, and $w=1$ for the original Hipparcos distance.   $\sigma_{\bar{x}}$ is the standard error, $\sigma$ is the standard deviation, and $w$ is the weight.  Lastly, the absolute Wesenheit magnitude ($W_{VI_c}$) for SU Cas implied by that distance should be corrected for contamination from the companion \citep[][see also \citealt{tu12}]{ev91,ev92}.

\section{{\rm \footnotesize CONCLUSION \& FUTURE RESEARCH}}
New XMM-Newton and OMM $JHK_s$ photometry for Alessi 95, in conjunction with existing UCAC3 and 2MASS observations, imply cluster parameters of: $E(J-H)=0.08\pm0.02$ and $d=405\pm15$ pc (Figs.~\ref{fig-pm},~\ref{fig-cmd}).  The determination confirms the findings by \citet{tu12}, and in particular that SU Cas is an overtone pulsator \citep{gi82,ev91}.  Distance estimates for SU Cas (i.e., ISB, Hipparcos, cluster membership) converge upon $d=414\pm 5(\sigma_{\bar{x}}) \pm 10 (\sigma )$ pc (\S \ref{s-dsucas}).  The small uncertainty places SU Cas in a select group of classical Cepheid calibrators ($\delta$ Cep and $\zeta$ Gem) that exhibit precise distances owing to cluster membership and the availability of trigonometric parallaxes.  Yet the establishment of an HST parallax for SU Cas remains desirable in order to corroborate (or \emph{refute}) the results.  HST FGS images are available for SU Cas and were obtained as part of proposal ID 10113 \citep{bon04}. Moreover, reducing the UCAC3 proper motion uncertainties and potentially extending the analysis to fainter stars would strengthen the distance derived.  Precise proper motions for fainter stars near SU Cas may be obtained from photographic plates stored at the CfA \citep[][DASCH]{gr07}.\footnote{Digital Access to a Sky Century @ Harvard (DASCH), \url{http://hea-www.harvard.edu/DASCH/}}  The plates offer unprecedented multi-epoch coverage spanning a $\sim100$ year baseline, and uncertainties are further mitigated owing to sizable statistics ($\sim (5-10) \times 10^2$ plates per object).  

Lastly, XMM-Newton X-ray data proved pertinent for segregating cluster members from field stars, thereby highlighting a novel approach for establishing improved distances for cluster Cepheids \citep[see also][]{ev11}.   Identifying members of Cepheid clusters using X-ray and proper motion data, as demonstrated here for SU Cas, may invariably bolster the Cepheid calibration \citep[e.g.,][]{ng12} and efforts to establish extragalactic distances \citep[e.g.,][]{ss10,ge11,fm10}.  Admittedly, considerable work remains to fine-tune the former approach and maximize full use of the available X-ray data.  Furthermore, additional work is needed to anchor the long-period end of the Galactic Cepheid calibration, despite continued advancements to secure the short and intermediate-period regimes.  Future research shall aim to assess the viability of establishing long-period Cepheid calibrators from their membership in spiral arms, which are likewise delineated by young open clusters \citep[see][]{ma11c}.

\subsection*{{\rm \scriptsize ACKNOWLEDGEMENTS}}
\scriptsize{DM is grateful to the following individuals and consortia whose efforts lie at the foundation of the research: OMM (E. Artigau, R. Lamontagne, R. Doyon), UCAC3 (N. Zacharias), 2MASS, N. Evans, XMM-Newton (N. Loiseau, R. G. Riestra), E. Guinan, S. Engle, F. van Leeuwen/M. Perryman (Hipparcos), P. Stetson (DAOPHOT), WEBDA (E. Paunzen), DAML (W. Dias), CDS, arXiv, and NASA ADS.  WG is grateful for support from the Chilean Center for Astrophysics FONDAP 15010003 and the BASAL Centro de Astrofisica y Tecnologias Afines (CATA) PFB-06/2007.}

\end{document}